\documentclass[useAMS,usenatbib,usegraphicx]{mn2e}  
\usepackage[usenames]{color}

\newcommand{\be}{\begin{equation}}
\newcommand{\ee}{\end{equation}}
\newcommand{\bea}{\begin{eqnarray}}
\newcommand{\eea}{\end{eqnarray}}

\newcommand{\LE}{\it LensExplorer\ \rm}

\title[Lens Explorer] 
{LensExplorer, a tool for the visualization of the gravitational lensing effect 
 in the Hubble Frontiers Fields clusters}  

\author[J.M. Diego]  
  {Jose M. Diego\footnote{jdiego@ifca.unican.es} \\ 
  IFCA, Instituto de F\'isica de Cantabria (UC-CSIC), Av. de Los Castros 
s/n, 39005 Santander, Spain\\}

\date{Draft version \today}  
  
\pagerange{\pageref{firstpage}--\pageref{lastpage}}  
  
\begin{document}  
\maketitle  
 
\label{firstpage}  
\begin{abstract}  
\LE is an IDL GUI code that can be used to visualize the lensing deflection field derived in the Hubble Frontier  
Fields clusters. The code can be used for both, research and educational purposes as it offers and intuitive and 
user friendly way to explore the solutions in the different clusters. \LE can be used 
to predict the magnifications at different redshifts of background sources and for a given location. 
We present in this document a quick guide to get started with \LE. 
The code and lens models for the cluster A2744 and MACSJ0416.1-2403 are 
publicly available from this website http://www.ifca.unican.es/users/jdiego/LensExplorer/ 
where the new solutions (and new versions of the code) will be also uploaded as they become available.
\end{abstract}  

\section{Introduction}\label{sect_intro}  
Gravitational lensing in galaxy clusters is quickly becoming a precision tool to determine the 
dark matter distribution in clusters thanks to the large number of multiple images that are routinely 
being discovered with deep observations, primarily from the Hubble telescope. An early example of the potential 
of deep Hubble observations in clusters was introduced in \cite{Broadhurst05a} where hundreds of lensed images 
where observed for the first time on a single cluster (Abell 1689). More recently, the CLASH program 
\cite{Postman2012b} has unveiled a wealth of multiply lensed galaxies around galaxy clusters also from Hubble 
observations. However, the observations in the CLASH program, did not take full advantage of the capacity 
of the Hubble telescope to make extremely deep observations. 
In the latest attempt to push for the limits of the Hubble telescope, the ongoing Hubble Frontiers Fields (HFF) 
program\footnote{http://www.stsci.edu/hst/campaigns/frontier-fields/}
promises to deliver a legacy of observations that will bring this field to a new standard. 

The HFF clusters are particularly interesting for several reasons. First, they are merger clusters with 
strong interactions between the different components. In some cases, the strength of the interaction can 
be quantified by combining observations of the lensing effect (that are used to derive the matter distribution) 
and X-rays that trace the plasma. Often these observations are complemented with radio observations that can reveal 
regions with shocks. In the near future, these observations will be complemented by yet another important observable, 
the Sunyaev-Zeldovich effect, whose technology is reaching the level to be competitive with the previous ones. The 
Sunyaev-Zeldovich effect together with the X-rays and radio observations, contribute with additional information 
about the dynamical state of the cluster. When complemented with the lensing studies, a multiwavelength study can reveal 
some of the properties of the DM as it has been already shown in some cases of merging clusters. The HFF clusters will 
play a fundamental role in the near future to understand the process of structure formation, since these clusters 
are captured in the process of merging, where the least understood physical mechanisms involved in structure 
formation can be studied in more detail. But also, the HFF will play a key role in understanding the nature of 
dark matter since the merger of clusters offers a unique opportunity to study the self-interaction properties of 
the dark matter. 

On a different level, the fact that these clusters are morphologically disturbed interesting consequences that 
make them ideal targets for other interesting studies. 
The elongation of the HFF clusters result in larger than normal areas 
where the magnification is significantly increased. Observations in other wavelengths 
can take advantage of the natural magnifying power of these lenses and reach further. The large magnifying area of 
the HFF clusters imply that these clusters offer the best chance to have a highly magnified source behind the cluster and 
hence the possibility to observe the faintest sources. Studies based on the HFF are opening a door to directly constrain 
the epoch of reionization by exposing the population of first galaxies that emerge from this period. The HFF program can 
be hence seen as the Hubble telescope looking at the furthest regions of the Universe through the most powerful natural 
telescopes (or lenses) known.

For the purpose of planning and understanding the observations it is important to have ways in which the 
magnification in a given position and for a given redshift can be estimated. It is also important to consider that 
observers in other fields may not be familiar working with magnification or deflection maps from gravitational 
lensing so having access to a user friendly interface might become a welcomed resource. The spirit behind \LE is 
to facilitate the use of lensing products by members of the community that are not directly involved in gravitational 
lensing analysis. \LE offers the possibility to easily estimate the magnification in the desired region to 
people with little or no experience with gravitational lensing data. This is achieved through a Graphic User 
Interface (GUI) that is fast, easy to use, intuitive, and user friendly. 

An additional goal of \LE is to be used as a tool to identify new systems. New systems may 
still be identified in the data that a patient and skilled user may be able to unveil using \LE. 

Finally, although the primary goal of the code is to be used as a scientific tool, \LE can be exploited 
also as a powerful application for educational purposes. 
Students that are learning about gravitational lensing will quickly 
develop an intuition about how gravitational lenses work. Moreover, they will be using real models derived from 
real data instead of working with idealized models and data sets. 

The current version of \LE has two of the HFF clusters implemented (A2744 and MACSJ0416). The remaining 
clusters will be implemented as the data and lens models become available. The models implemented in this release 
are based on the results of \cite{Lam2014} for A2744 and \cite{Diego2014b} for MACSJ0416. The model for A2744 
presented in this release is a simplified version of the one presented in \cite{Lam2014} so small differences are 
expected when comparing with the results presented in \cite{Lam2014}. A more detailed model 
will be included when the full data set for A2744 is made available and a new improved analysis is completed. 
For the case of MACSJ0416, the solution accompanying this release corresponds to the case (iii) described in 
\cite{Diego2014b}. Also, and as in the case of A2744, this solution will be updated when the full data for this 
cluster is re-analized. 
In both cases, the results are based on the free-form method WSLAP+ described (and applied) in more detailed in 
the following papers, \cite{Diego05a,Diego05b,Diego2007,Ponente2011,Sendra2014,Diego2014a,Diego2014b}, 
and \cite{Lam2014}. 

\LE will be updated on the official website\footnote{http://www.ifca.unican.es/users/jdiego/LensExplorer/} 
each time a new cluster is implemented or a current model is improved after a re-analysis of new HFF data. 
Additional clusters like A1689 or the CLASH clusters may be also implemented together with future versions/releases 
of the code/models. Additional models derived by independent teams and using different techniques can be found in 
the HFF webpage.\footnote{http://archive.stsci.edu/prepds/frontier/lensmodels/}

\section{Using LensExplorer}

\begin{figure*}  
\centerline{ \includegraphics[width= 18cm]{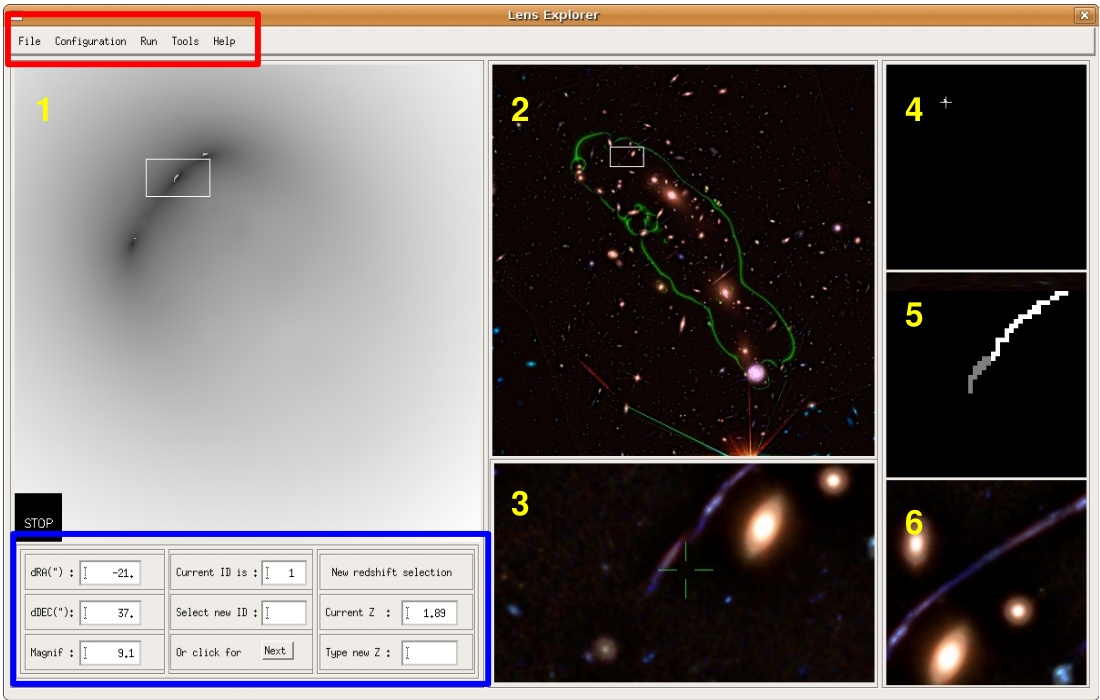}}          
   \caption{\LE at work in {\it Explore} mode. 
            The GUI is divided into 6 display windows (marked with yellow numbers 1-6), 
            an interactive information box (marked with a blue rectangle in the bottom-left) and a menu 
            bar (marked with a red rectangle on the top-left). 
            }
   \label{fig1}  
\end{figure*}  

In this section I will assume that you have a working version of IDL installed in your computer 
(or that you have acces toa machine with an IDL license) but that you 
have no knowledge of the IDL language. If this is your case, I explicitly mention in parenthesis the very few 
commands you need to type in IDL in order to compile and run the code.  
If, on the other hand, your knowledge of IDL is more advanced, 
you may want to check for ways to improve the code and maybe contribute to improvements 
to future versions of the code, which would be very much appreciated.

I also assume that your knowledge of gravitational lensing is limited. Because of this, I include below a few useful 
definitions that will appear through the text. If you are familiar with this terms, you can safely ignore this part.

$\bullet$ {\it Image plane}. In gravitational lensing, this refers to the virtual plane that is perpendicular 
to the line of sight ad at the same redshift of the galaxy cluster (or lens).

$\bullet$ {\it Source plane}. This is a similar plane but at the redshift of the background source (behind the lens) 
and that is being lensed. 

$\bullet$ {\it Delens}. When you see this expression it refers to raytracing back the image (photons or cursor position) 
from the image plane to the source plane (backwards projection). 
A delensed image should in theory look like the original background image (if gravitational lensing never happened). 

$\bullet$ {\it Relens}. This is the opposite of {\it Delens} and corresponds to taking the photons from 
the source plane, through the lens and forming the lensed image(s) in the image plane (forward projection).  
As the lensing problem is intrinsically non-linear in nature, that means that during 
the forward projection there may be different regions in the image plane where the forward projection may land. 
These are the positions where the multiple images are expected.

$\bullet$ {\it Deflection field}. The angle at which the photons are bended when passing through the lens. 
It can be viewed as two 2D-images with the same dimension of the image in the image plane. One image represents the 
deflection in the horizontal axis and the other the deflection in the vertical axis.  
 
$\bullet$ {\it Magnification}. In theory, the magnifying factor that the lens is applying over the background galaxy 
(source). The magnification can be viewed in terms of area or flux (factor by which the area of the source 
or flux is being magnified). 
In practice, this idealized number is just an approximation to the real, effective magnification specially 
for large magnifications. For instance, if the model predicts a magnification of 100 in a given position, most likely 
a background galaxy that is lensed forming an image at that position will appear as 10, maybe 20 (at most) times larger 
in area than the original image. The same argument that applies to the observed area applies to the observed flux. 
The real magnification of the background source depends on its morphology and precise location/orientation with 
respect to the caustics. The magnification predicted by the lens model should be taken as an approximation, specially 
in the regions with very high magnifications. Small changes in the lens model or background source position (or redshift) 
can result in big changes in the magnification. 

\subsection{Ready, Set, Go!}
When downloaded, the uncompressed package consists of a single self-contained IDL routine and a series 
of sub-directories containing the data and lensing models. 
After compiling the code in IDL ({\bf IDL$>$.r LensExplorer}), the routine {\it go} will launch the application 
({\bf IDL$>$go}). 
A small screen will prompt you to select one data set (or cluster). After you select on of the data sets, you will 
see a screen like the one shown in figure \ref{fig1} except that all screens and data values will be blank.  
To run the main part of the code you can do so by clicking with the mouse in the {\it Start} option under the 
menu option {\it Run}. 
At this point, the code reads the data, a splash screen briefly shows the strong lensing data set (in Window 1) 
used for the mass reconstruction and soon after the code enters in {\it Explore} mode (see figure \ref{fig1}). 

The GUI is divided in three parts. There are 6 {\bf graphical windows} (marked 1-6 in figure \ref{fig1}) 
that display 2D information about the lens and source planes. 
In the bottom-left part of the GUI (marked with a blue rectangle in figure \ref{fig1}), 
the {\bf information panel} gives the user useful information. 
Some of this information can be updated interactively. In the top-left (marked with a red rectangle in figure 
\ref{fig1}), a {\bf menu bar} contains some useful tools and actions that control the application. 
These three parts are described in more detail below.

\subsection{Graphical windows}
The graphical windows are shown in figure \ref{fig1} and are marked from 1 to 6 (in yellow). 

$\bullet$ Window 1 is the main device to interact with the application. 
It shows the distance of the image plane to the delensed position of the cursor when the entire image plane is delensed. 
In other words, the dark-grey regions in window 1 mark the points where counter-images are expected 
for the given cursor position and redshift. 
When clicking with the mouse in Window 1, the code projects (delens) the cursor position 
back to the source plane located at the redshift displayed in the {\it Current Z} box of the information panel. 
This delensed position in the source plane is shown as a cross in Window 4. 
Window 1 also displays the data points for a given system (in light grey). The delensed positions of these points 
are also displayed in Window 4. When clicking with the mouse in Window 1, a white rectangle around the mouse position 
marks the region that is displayed in Window 3 at full resolution. This window corresponds to a region of 
$12.6\times7$ arcseconds$^2$ when the field of view is 2.4 arcminutes across (and scales accordingly when the field of 
view changes). The same region is shown in the context of the entire field of view in Window 2. 

$\bullet$ Window 2 displays the Hubble image with the critical curves superimposed. It also shows the rectangular region 
being zoomed in Window 3. When working on {\it DeRe-lens} mode (see section \ref{sect_tools}) Window 2 displays 
the relensed image and can be explored with the mouse. In this case, a small region of $6.6\times6.6$ arcseconds$^2$ 
around the mouse position (for a field of view is 2.4 arcminutes) 
is interactively displayed in Windows 6 (relensed model) and 7 (original data). 

$\bullet$ Window 3 shows the Hubble data at full resolution in the rectangular region shown in Windows 1 and 2. 
This window is also used to select the image being relensed-delensed as described in section \ref{sect_tools} below.
The Hubble image has been processed to reduce the diffuse light. 

$\bullet$ Window 4 shows the source plane. In {\it Explore} mode, it shows the delensed arcs for the current system and 
the delensed position selected by the mouse (cross). In  {\it DeRe-lens} mode it shows the delensed image that is 
being selected to be relensed. 

$\bullet$ Window 5, when in  {\it Explore} mode it shows the data points (arcs or observed data or constraints) 
used to reconstruct the mass. 
Each arc is coded with a different index running from 1 to N, with N the total number of systems. 
When in {\it DeRe-lens} mode, 
this window shows a zoomed version (x2) of the source in the source plane and also a full resolution version of the 
relensed image as predicted by the model. 

$\bullet$ Window 6 is active only when in  {\it DeRe-lens} mode and it shows the original Hubble data at the 
same position and with the same scale as in Window 5. 

\begin{figure*}  
\centerline{ \includegraphics[width= 18cm]{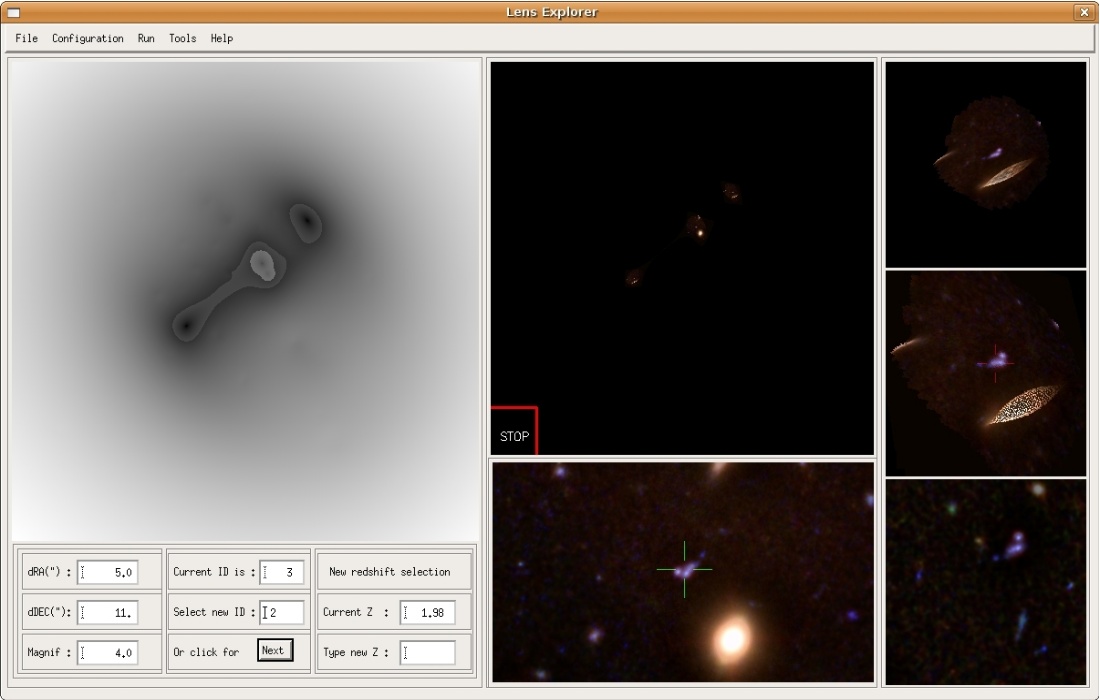}}          
   \caption{\LE in {\it DeRe-lens} mode. 
            Some Windows change their role in this mode. Window 2 becomes now active to the mouse and 
            incorporates its own STOP function to leave this mode and return to the {\it Explore} mode.    
            }
   \label{fig2}  
\end{figure*}  

\subsection{Information Panel}
Located in the bottom-left of the GUI (and marked with a blue rectangle in figure \ref{fig1}), this panel 
contains useful information, some of which can be updated by the user.
The left column shows the relative distance (in arcseconds) from the centre of the field of view. Keep in mind that 
these distances assume a flat, planar geometry so in order to translate into real RA and DEC you need to take into 
account the factor cosine(DEC). Below RA and DEC, the code shows the magnification under the current position of the 
mouse (click) and for the fiducial redshift. 
The fiducial redshift is taken as z=3 by default and is the same redshift that is used to 
compute the critical curve shown in Window 2. If you want the magnification (and deflection filed) for a different 
redshift (of the source) you can recompute the magnification (and critical curves) by using the 
tool {\it Magnification(z)}. See section \ref{sect_tools} for details. 

The middle column in this panel shows information about the current system. This column is useful to know what system is 
being displayed in Windows 1 and 4 but it can also used to jump to a particular system. To do so, if you are in 
{\it Explore} mode you need to leave this mode by clicking in the {\it STOP} black square of Window 1. 
If you forget to click {\it STOP} in this window the program will give you an error but will still continue running. 
At any point, or if the program is not responding due to some errors, you can always reselect again the {\it Explore} 
mode from the {\it Tools} menu.  
Then you can type the new ID in the box {\it Select New ID} and type enter. Alternatively you can jump from one 
system to the next by clicking on the box {\it Next} at the bottom of this column. 
The right column of the information panel contains useful information about the redshift of the current system. The value 
displayed in {\it Current Z} will be taken as the new fiducial value every time you recompute the magnification. 
Every time the magnification is computed, the critical curves change accordingly to the new redshift. 
The redshift of a given system (or the fiducial redshift for the new magnification and critical curves) 
can be modified by the user at any time by typing a new value in the box {\it Type new Z} and pressing enter. 

\subsection{Menu bar}
The menu bar is located in the top-left of the GUI (and marked with a red rectangle in figure \ref{fig1}. Some of the 
functions are not 100\% functional yet (Version 1.0) and will be updated in future versions of the code but they are 
included here for completeness purposes. The menu bar contains 5 items; {\it File, Configuration, Run, Tools} 
and {\it Help}.

$\bullet$ {\it File}. This menu element contains three options; {\it Fast Load, Select Files} and {\it Exit}. The option 
{\it Fast Load} brings in the window that allows to select among the different data sets. This window launches 
automatically every time the program is run for the first time in an IDL session. The option {\it Select Files} allows the 
user to check the data files and models that are being loaded by the program. It also allows to change these files. 
This second possibility will be more useful when different models (for the same cluster) are integrated with new releases 
of the code. The first version of the code comes with just one set of data and models per cluster. Finally, the 
option {\it Exit} quits the program and returns to the IDL command line. 

$\bullet$ {\it Configuration}. This element contains only one option ({\it Set Parameters}) that let's the user check 
(and change) some parameters including the cosmological model and redshifts of the lens and default fiducial. 
These parameters are pre-loaded fro each cluster when selecting a particular data set.

$\bullet$ {\it Run}. Also containing one element ({\it Start}), loads the data and enters {\it Explore} mode 
starting with system ID=1. 

$\bullet$ {\it Tools}. This element contains different options that are described in more detail in 
section \label{sect_tools}.

$\bullet$ {\it Help}. Contains different elements not fully implemented in the current release about the version 
of the code and different help options.

\section{Menu and Tools}\label{sect_tools}
The {\it Tools} item in the menu option has 5 different options; {\it Explore,DeRe-lens,Magnification(z), Check ID} 
and {\it Take Snapshot}. 

$\bullet$ {\it Explore}. This is probably the feature that will be used most often. It allows to explore the lens 
  plane with the mouse on Window 1. By moving the cursor around Window 1, the code recomputes the predicted position 
  of the counter-images for an alleged image at the position of cursor and for the redshift shown in {\it Current Z} in 
  the {\it information panel}. The predicted positions are shown as dark-grey with darker grey marking the most likely 
  position for the counter-images. 
  The Hubble image around the cursor position is shown at full resolution in Window 3 while 
  Window 2 shows the rectangular area that is being displayed in Window 3. To exit the {\it Explore} mode and to update 
  some of the values of the {\it information panel} (ID and z) or to choose a different {\it Tool}, click in the 
  bottom-left corner of Window 1 ({\it STOP}). 

$\bullet$ {\it DeRe-lens}. This option allows the user to select a source to be delensed into the source plane and 
   relensed back into the image plane. The source to be delensed is selected while in the {\it Explore} mode. Move the 
   cursor until the source you want to select is at the centre of Window 3. Click {\it STOP} at the bottom-left 
   corner of Window 1 and select {\it DeRe-lens} from the {\it Tools} menu. The code will select an area to delens 
   around your selected position (medium grey in Window 1) and preselect an area in the image plane where to look 
   for counter-images (dark grey in Window 1). The search for counter-images normally takes 10-15 seconds so be patient. 
   The delensed source will be shown in Window 4 at ACS equivalent resolution and and Window 5 zoomed by a factor 2. 
   When the relensed map is finalized it will be shown in Window 2. You can now explore Window 2 with the mouse in a 
   similar way as you explore Window 1 when in {\it Explore mode}. An area of $6.6\times6.6$ arcseconds$^2$ around 
   your selected position will be shown in Window 5 for the relensed model image and Window 6 for the original Hubble 
   data. Both the scale and centre of the images in Windows 5 and 6 are the same. 
   To exit  {\it DeRe-lens} mode click on the bottom-left corner of Window 2 ({\it STOP}) and you will return to 
   the {\it Explore} mode.

$\bullet$ {\it Magnification(z)}. Recomputes the magnification map and critical curves for the redshift in use and 
that is being displayed in the field {\it Current Z} of the {\it information panel}. To compute the magnification 
(or critical curve) at a desired redshift the user needs first to type the new redshift in the {\it information panel}. 
The critical curve is updated in Window 2 when returning to the {\it Explore} mode. 

$\bullet$ {\it Check ID}. It allows the user to check what systems are being displayed in Window 5. This option is 
  useful when trying to identify new candidates and to make sure that the candidate is not already included in the 
data set. Also useful to remind the user what system is near the current system that is being displayed in Window 1. 
Right click with the mouse on this window (5) to exit this mode.

$\bullet$ {\it Take Snapshot}. Not implemented yet in the current version of the code but it will let the user 
  select a window to create a snapshot of a selected graphic window and store it in jpg format. 

\section{Acknowledgments}  
If you found this code useful for your research, or educational project the author would be grateful if a 
proper reference to this paper is included in your work. JMD acknowledges the hospitality of the Physics and 
Astronomy department at UPenn during the time in which this code was being developed. 
JMD acknowledges support of the consolider project CAD2010-00064 and 
AYA2012-39475-C02-01 funded by the Ministerio de Economia y Competitividad. 
The data used in the lensing reconstruction were derived as part of the HST Frontier Fields program 
conducted by STScI and were obtained from the Mikulski Archive for Space Telescopes (MAST). The author would like 
to thank this program and the people that made it possible and available for the community. 


\bsp  
\label{lastpage}

\bibliographystyle{mn2e}
\bibliography{MyBiblio} 

\end{document}